\documentclass[referee]{raa}            

\usepackage[utf8]{inputenc}

\usepackage{graphicx,times}             
\usepackage{grffile}

\usepackage{natbib}
\usepackage{amssymb,amsmath}
\bibpunct{(}{)}{;}{a}{}{,}
\usepackage{multirow}
\usepackage{ae,aecompl}
\usepackage{fourier}
\usepackage{graphbox}

\usepackage{latexsym}	
\usepackage{gensymb} 
\usepackage{color}

\usepackage{ulem}

\usepackage[pagebackref=true]{hyperref}

\renewcommand{\vec}[1]{\boldsymbol{#1}}

\begin{document}

\title{Measurement of the Primary Beam of the Tianlai Cylindrical Antenna Using an Unmanned Aerial Vehicle}

   \volnopage{Vol.0 (20xx) No.0, 000--000}      
   \setcounter{page}{1}          

    \author{Jixia Li 
      \inst{1,2}
    \and Nanben Suo
      \inst{1}
    \and Shenzhe Xu
      \inst{1,2}
    \and Shijie Sun
      \inst{1,2}
    \and Shifan Zuo
      \inst{1,2}
    \and Yougang Wang
      \inst{1,2,3}
    \and Fengquan Wu
      \inst{1,2}
    \and Juyong Zhang
    \inst{4}
    \and Peter Timbie
    \inst{5}
    \and Reza Ansari 
    \inst{6}
    \and Albert Stebbins
    \inst{7}
    \and Xuelei Chen
      \inst{1,2,3}
   }

   \institute{
        National Astronomical Observatories, Chinese Academy of Sciences, Beijing 100101, China; {\it xuelei@cosmology.bao.ac.cn}\\
        \and
        State Key Laboratory of Radio Astronomy and Technology,  Beijing, 100101, China\\
        \and
        School of Astronomy and Space Science, University of Chinese Academy of Sciences, Beijing 100049, China\\
        \and 
        Hangzhou Dianzi University, Hangzhou 310018, China\\
        \and
        Department of Physics, University of Wisconsin – Madison, Madison WI 53703, USA\\
        \and
        Universit\'e Paris-Saclay, CEA/Irfu/DAp, Orme des Merisiers, 91191 Gif sur Yvette, France\\
        \and 
        Fermi National Accelerator Laboratory, P.O. Box 500, 
        Batavia IL 60510, USA\\
   \vs\no
   {\small Received 20xx month day; accepted 20xx month day}}

 \abstract{The Tianlai Cylinder Pathfinder Array consists of three adjacent cylindrical reflectors fixed on the ground, each 40 meters long and 15 meters wide, with the cylinder axis oriented along the North-South (N-S)direction.  Dual linear polarisation feeds are distributed along the focus line, parallel to the cylinder axis. Measurement of the primary beam profile of these cylindrical reflectors is difficult, as they are too large to be placed in an anechoic chamber. While the beam profile along the East-West (E-W) direction can be measured with the transit observations of bright astronomical radio sources, the beam profile along the N-S direction remains very uncertain. We present a preliminary measurement of the beam profile of the Tianlai cylindrical antenna along both the N-S direction and E-W direction in the frequency range of 700-800 MHz, using a calibrator source carried by an unmanned aerial vehicle (UAV) flying in the far field. The beam profile of the Tianlai cylindrical antenna is determined from the analysis of the auto-correlation signals from the  the cylinder array correlator, taking into account the emitter antenna beam profile, itself measured with a dipole antenna on the ground.
The accuracy of the  UAV-based determination of the cylinder beam profiles is validated by comparing the results with the one derived from bright astronomical source transits, and with simulated beams.  
}

   \authorrunning{Jixia Li et al. 2025}            
   \titlerunning{Measurement of Tianlai Cylinder beam using UAV}  

   \maketitle


\section{Introduction}
\label{sec:intro}

The Tianlai Cylinder Pathfinder Array (TCPA) is a radio interferometer array designed to test techniques for 21~cm intensity mapping \citep{Chenxl2012tianlai}, which has the potential of measuring the equation of state of dark energy from the baryon acoustic oscillation (BAO) or constraining inflation models \citep{Reza2012LSS,Xuyd2016inflation}.
Like many other similar intensity mapping (IM) or Epoch of Reionization (EoR) experiments, such as CHIME \citep{CHIME2022}, HIRAX \citep{HIRAX2016}, LOFAR \citep{LOFAR2016}, MWA \citep{MWA2013}, HERA \citep{HERA2017}, TCPA  is an interferometer array with a large number of receivers to achieve a fast mapping speed. These experiments often employ a very large number of array elements observing in drift-scan mode.  To reduce the cost, array elements are typically fixed or have
adjustable pointing in declination only. These interferometers with fixed reflectors pose a challenge for the calibration measurement of the beam pattern using bright celestial sources. 

The beam pattern of a radio telescope is usually measured using  well-known bright radio point sources as calibrators, because the antenna is too large to be fitted into an anechoic chamber. But, for the many HI intensity mapping or EoR experiments with non-steerable antennas, the measurement can only be performed along the East-West (E-W) direction when the bright point sources transit through the beam. As the number of bright point sources available for such measurement is very limited, this can only be performed at a few fixed zenith angles or, equivalently, declination angles, which are too sparse to derive the beam pattern in the North-South (N-S) direction. 

The Tianlai cylinder array, which consists of three fixed cylinders aligned in the N-S direction, also faces this problem. In order to reconstruct the sky from its observation data, it is necessary to know the primary beam pattern of the cylinder \citep{Zuosf2019eigen,Zuosf2021tlpipe}. The main lobe of the primary beam of the cylinder reflector is a narrow strip along the North-South direction passing through the zenith. The cylinder form of the antenna suggests a natural multiplicative  decomposition of the beam pattern along the E-W and N-S direction. Along the E-W direction, the pattern can be measured by observing the transit of bright point sources, which pass through the narrow strip of the primary beam, and shows up as a clear rise and fall of its  signal strength \cite{Lijx2020tianlai}. However, measuring the N-S direction pattern is much more difficult. Previously, \citet{Sunsj2022sim} has simulated the beam pattern for a single Tianlai cylinder with the feed array, and we have used an analytical beam model fitted to this result in the simulations of the Tianlai array imaging \citep{Yu2023,YuKF2024} and beam-forming \citep{YuZJ2024}. However, it is difficult to assess the accuracy of the beam model without actual measurement. 

Some methods have been proposed to measure the beam pattern for cylinder telescopes, such as holographic beam mapping \citep{CHIME2016holo,Thompson2017book,ATA2011holo}, or using the Sun as a calibrator \citep{CHIME2022sunbeam}. However, to use the holographic beam mapping method, a calibrated, large, steerable dish antenna is needed for cross-correlation, which is not available in many cases, including ours. The Sun calibrator method can only be used to measure the declination range of the ecliptic circle, i.e. $-23.5^\circ < \delta <23.5^\circ$, and the solar emission itself changes all the time, which affects the precision of the measurement. 

In recent years, beam measurement using a calibrator source carried by unmanned aerial vehicles (hereafter called drones) has proven to be a promising technique based on its flexibility and cost-effectiveness  \citep{Chang2015beamcali,Zhangjy2021dishbeam,CHIME2024drone,TONE2025drone}. For large, fixed antennas that do not fit into an anechoic chamber and can not employ the holographic mapping method, this is practically the only feasible method at present. By flying over the antenna array in the far field, one can directly measure the beam pattern, not only for the main lobe of the beam but also for the side lobes. 

However, to measure the beam pattern directly requires the source to be located in the far field zone of the antenna. The distance $d$ of the far field is estimated to be 
\begin{equation}
    d=\frac{2 D^2}{\lambda},
\end{equation}
where $D$ is the aperture of the antenna and $\lambda$ the wavelength. This corresponds to $d=1125 \mathrm{\,m}$ 
for $D=15 \mathrm{\,m}$ and $\lambda=0.4 \mathrm{\,m}$, for example, which will then
 require the drone to fly well above the usual height limit of a few hundred meters. 
 Although it is possible, in principle, to derive the far-field beam patterns from near-field measurements, this would require measurements on a very fine grid and accurate phase measurements, which are very challenging to perform in the field. In this paper, we present our measurement of the beam pattern for the Tianlai Cylinder by a drone.  In our measurement, we have flown our drone at a relative height of 1220~m. 

This paper is organized as follows. In Section \ref{sec:setup}, we describe the experiment setup, including the Cylinder Array, the drone, and the calibrator source it carries, the flight plan and the data calibration method. In Section \ref{sec:drone_cali}, we describe the calibration of the beam pattern of the calibrator source antenna, which is needed for correcting the cylinder beam measurement. Finally, the results of the cylinder beam pattern measurement are presented in Section \ref{sec:results}. We conclude in Section \ref{sec:discussion}.

\section{The Experiment Setup}
\label{sec:setup}

As noted in the Introduction, the TCPA is fixed on the ground. We fly our drone above the array, carrying an emitter, which produce a bandpass-filtered noise signal which is used as the calibration signal. The TCPA correlator system is operating in the meantime, which records the signal produced by the drone along with the sky signal and receiver noise. As the drone flies to different positions relative to the array, the variation of the its signal strength is measured, and the beam pattern  of the cylinder array is derived. 
Below we present the details of this experiment. 

\subsection{The Tianlai Cylinder Pathfinder Array}
\label{subsec:cylinder_array}

\begin{figure}[ht]
    \centering
    \includegraphics[width=0.6\linewidth]{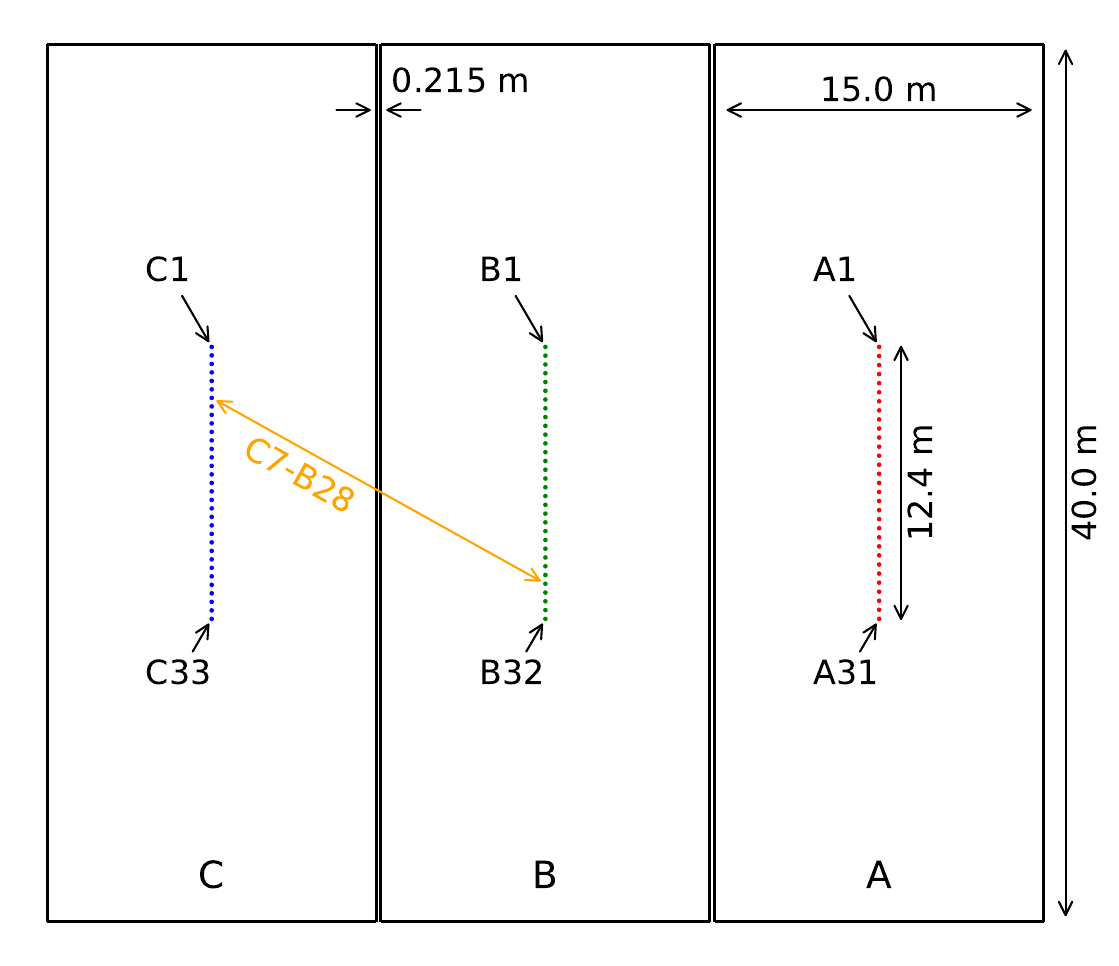}
    \caption{The Tianlai Cylinder Pathfinder Array and the designation of its feeds. The cylinders are aligned in the N-S direction, with a gap of 0.215 m between adjacent ones. 
}
    \label{fig:array_config}
\end{figure}

The TCPA consists of three adjacent parabolic cylinder reflectors fixed on the ground, each 40 m × 15 m, with their axes oriented in the N-S direction. Dual-polarization dipole feeds are placed along the focus line of each cylinder \citep{Chenzhp2016design}. The cylinders are closely spaced in the E-W direction, as shown in Fig. \ref{fig:array_config} (reproduced from \citealt{Lijx2020tianlai}). 
The reflectors focus the incoming radio signal in the E-W direction, while allowing a wide field of view (FoV) in the N-S direction. At any moment its FoV is a narrow strip running from north to south through the zenith. The beam of the feeds limits the strip to about $\pm 60 \degree$ from zenith \citep{Sunsj2022sim,Cianciara2017feedsim}. 
As the Earth rotates, the beams scan the northern celestial hemisphere. The latitude of the telescope site is $44.15\degree$, so the FoV extends from $-16\degree$ declination up to $+90\degree$ and back down to $+76\degree$ on the other side.  
From east to west, the 3 cylinders are denoted as cylinder A, B, C, respectively, and equipped with 31, 32 and 33 feeds, respectively. From north to south, the feeds in each cylinder are labeled in numbers 1, 2, 3, etc. The northernmost feeds A1, B1, C1 are aligned, as are the southernmost feeds A31, B32, C33, and the distance between the northernmost and southernmost feeds is 12.4 m. The currently installed feeds occupy less than half of the cylinder length; the remaining space is reserved for additional feeds for future upgrades. The feed spacing for the A, B, C cylinders are slightly different: 41.33 cm, 40.00 cm and 38.75 cm, respectively, to reduce the grating lobe \citep{Zhangjiao2016sky}. The X and Y polarization denotes those along the N-S direction and E-W direction respectively. Each signal channel is designated by its cylinder, feed number, and polarization axis. For example, the E-W or Y polarisation signal of the 2nd feed in the middle cylinder will be noted as B2Y. 
The auto-correlations are denoted by the feed and polarization, e.g. C7 XX. The TCPA is designed to observe the sky between 400~MHz to 1500~MHz by changing the filters before the mixers \citep{Lijx2020tianlai}. Currently, filters centered on 750~MHz with a bandpass of 100~MHz are used, resulting in a frequency range of 700-800~MHz. 

\begin{figure}
\centering
\includegraphics[width=0.8\linewidth,align=c]{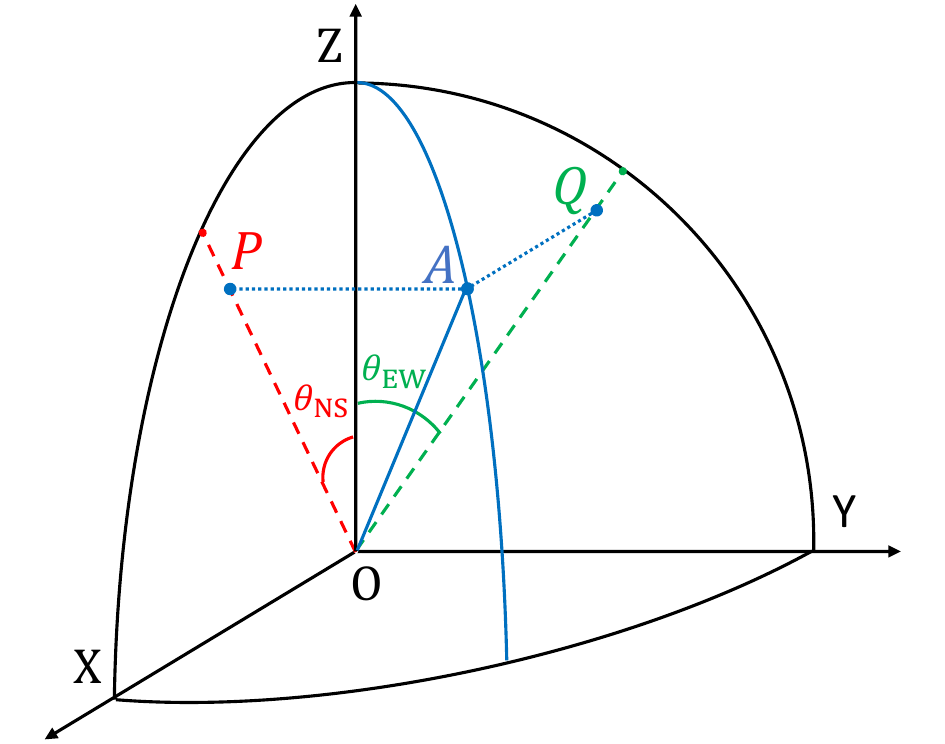}
\caption{The definition of projected angles $\theta_{\rm NS}$ and $\theta_{\rm EW}$. For an arbitrary direction A, P and Q are the vertical projection points on the XZ and YZ planes, respectively.  }
\label{fig:angle_def}
\end{figure}

We model the primary beam of the cylinder as a product of the NS and EW profiles,
\begin{equation}
B (\mathbf{n}) = B_{\rm NS}(\theta_{\rm NS}) \cdot B_{\rm EW}(\theta_{\rm EW})    
\label{eq:decomp}    
\end{equation}
where $\theta_{\rm NS}$ and $\theta_{\rm EW}$ denotes the angle with respect to the $Z$-axis for the projected vector on the $XZ$ and $YZ$ planes \footnote{Note here $\theta_{\rm NS}$ and $\theta_{\rm EW}$ denotes the pair of projected angles for arbitrary direction, while  
in \citet{Yu2023} the same symbols are used to denote the widths of the beam profile along the NS and EW directions.}, as shown in Fig.~\ref{fig:angle_def}.

\subsection{The Drone and Noise Source}

To measure the beam pattern of the cylinder reflectors, we use a calibrator noise source mounted on a drone to fly over the cylinder slowly in far field zone. The cylindrical antenna does not have an azimuthal symmetry and can not be characterised by a single value for its aperture. In our case, the aperture along the E-W direction is 15~m, while along the N-S direction the reflector has no focus; its equivalent aperture is determined by the illumination angle of the feed. So we use the E-W aperture to determine the far field distance. For the current observing frequency range 700-800 MHz, the {\em far field} distance is 1050-1200 m. In this measurement campaign, the drone was flown at a height of 1220~m above the central feed.

\begin{figure}
    \centering
    \includegraphics[width=0.7\linewidth]{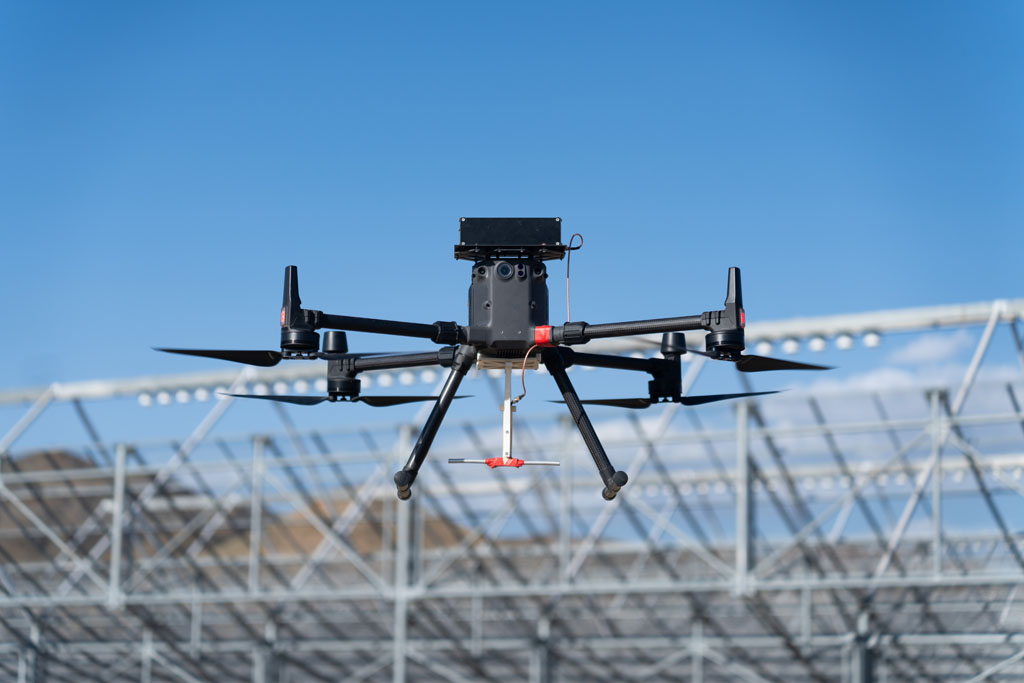}
    \caption{The DJI M300 RTK drone used in our experiment. The noise source box is mounted on the top of the drone. The dipole antenna is mounted below the drone and connected to the noise source emitter with a RF cable.}
    \label{fig:drone_photo}
\end{figure}

The drone used in our experiment is of type DJI M300 RTK, 
as shown in Fig. \ref{fig:drone_photo}. It is a commercially available drone; some relevant parameters are listed in Table \ref{tab:drone_params}. At the location of the TCPA, the ground elevation is about 1500~m above sea level. Starting from this site, the drone has a maximum endurance of about 55 minutes with empty payload and low level flight, but with payload, and especially to attain the ground height of over 1000 meters, the endurance is greatly reduced. The drone needs about 7 minutes to reach the height of 1220~m, the equivalent endurance can be reduced by as much as 15 minutes. For a safe return flight and touch down, a minimum of 12 minutes endurance should also be reserved. In our experiment, the weight of the whole payload carried by the drone is about 0.75~kg, corresponding to a maximum endurance of 43 minutes. Therefore, 
the available time for the actual measurement work is limited to about 15 minutes during each flight.

\begin{table*}
    \centering
    \caption{Parameters of the DJI M300 drone. (\url{https://www.dji.com/cn/support/product/matrice-300})}
    \label{tab:drone_params}
    \begin{tabular}{|l|l||l|l|}
    \hline
    Geometric Size (mm) & 810 ×670 ×430 & Max. take-off weight & 9 kg \\
    \hline 
    Horizontal hovering error  & $\pm0.1$ meter &
    Vertical hovering error  & $\pm0.1$ meter \\
    \hline
    Horizontal RTK locating error & 1 cm  &
    Vertical RTK locating error & 1.5 cm \\
    \hline
    Max. rising speed (S mode) & 6 m/s &
    Max. rising speed (P mode) & 5 m/s \\
    \hline
    Max. descending speed (S mode) & 5 m/s &
    Max. descending speed (P mode) & 4 m/s \\
    \hline
    Max. horizontal speed (S mode) & 23 m/s &
    Max. horizontal speed (P mode) & 17 m/s \\
    \hline
    Max. working elevation & 5000 m sea level &
    local ground elevation & 1500 m sea level \\
    \hline
    Max. tolerable wind speed & 12 m/s ($<$6 level) &
    Operating temperature & $-20^\circ\mathrm{C}$ to $50^\circ\mathrm{C}$ \\
    \hline
    Communication band 1 & 2.4-2.4835 GHz &
    Communication band 2 & 5.725-5.850 GHz \\
    \hline
    \end{tabular}
\end{table*}

\begin{figure}
    \centering
    \includegraphics[width=0.8\linewidth]{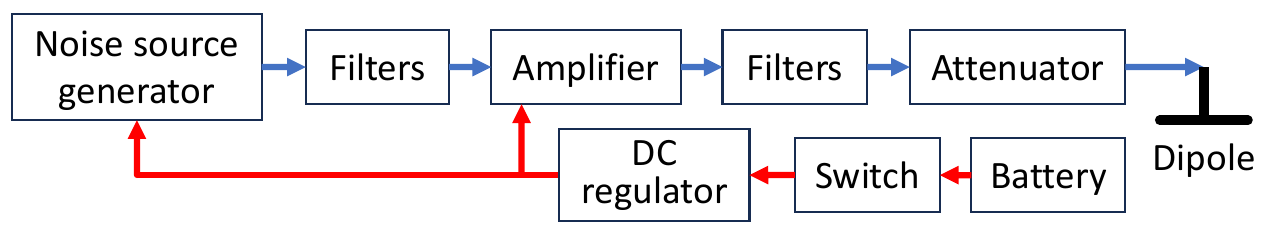}
    \caption{A block diagram of the noise source calibrator system. }
    \label{fig:noise_source_photo}
\end{figure}

The position of the drone is determined using the differential Global Positioning Systems (differential GPS). The drone has an embedded Real-Time Kinematic (RTK) GPS module to ensure $\sim 1 \mathrm{\,cm}$ location accuracy. The RTK base station is placed near the cylindrical antenna for operation. In the actual flight, if a target position is set, the drone will drift around it, resulting in a hovering error of $\pm 0.1\,$m (see Table \ref{tab:drone_params}). At a distance of 1200~m, it introduces an angle error of only $17''$, thus the positional error is negligible. However, under strong wind, the drone will tilt itself to counteract the wind, which affects the beam pointing of the dipole antenna. To avoid this problem, the measurements were made only under low wind speed conditions when the ground wind speed is less than 2 m/s.

The drone carries a noise emitter, which includes a dipole antenna mounted below the drone, and an electronics box mounted above the drone, connected by a radio frequency (RF) cable. The block diagram of the calibrator noise source system is shown in Fig.~\ref{fig:noise_source_photo}.
A noise source generator produces a wide band noise signal with a flat spectrum spanning 0.1~MHz to 1.5~GHz (1500~MHz). To increase the signal power, the noise is amplified by about 20~dB by a broad band RF amplifier (a Mini-Circuits ZX60-33LN-S+, with operating frequency range 0.5-3 GHz). A low-pass filter (Mini-Circuits VLF-800+) with ultra-flat $0.6 \pm 0.2~$dB passband ripple is inserted before the amplifier, followed by 710-850 ~MHz
bandpass filter (Mini-Circuits model VBFZ-780-S+), to suppress the out-of-band noise which may affect the operation control and location system of the drone. An attenuator is used to adjust the emitted signal level. 

\begin{figure}
    \centering
    \includegraphics[width=0.8\linewidth]{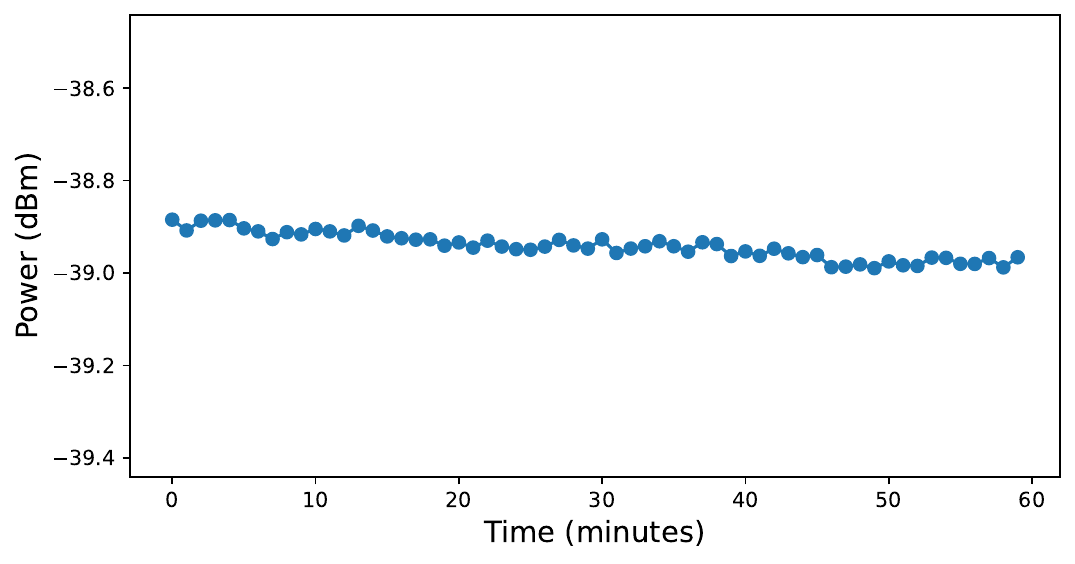}
    \caption{Stability measurement of the drone noise source system in 60 minutes.}
    \label{fig:dns_stability}
\end{figure}

We measured the stability of the noise source circuit over a period of 60 minutes with a spectrum analyzer, and estimate the noise power by averaging over the observational bandwidth. The result is shown in Fig. \ref{fig:dns_stability}. The time interval between two measurements is 1 minute. The variation during each measurement is within $\pm$0.02~dB, which is close to the precision of the spectrum analyzer. However, a steady decrease of about 0.1~dB can be observed for the whole duration of the measurement, probably due to the decreasing battery power. Within the 10 minute measurement time of the cylinder beam, the variation of the drone noise source signal can be ignored. 

The dipole antenna is made of an aluminum tube of diameter 7~mm and length 105~mm for each half of the dipole.
It is fixed to a bracket made by a 3D printer, which is the white part mounted below the drone in Fig. \ref{fig:drone_photo}. The dipole antenna resonates at 645~MHz, with a quarter-wave coaxial balun ensuring impedance match and current balancing.

\subsection{The Flight Plan}
\begin{figure}
    \centering    \includegraphics[width=0.8\linewidth,align=c]{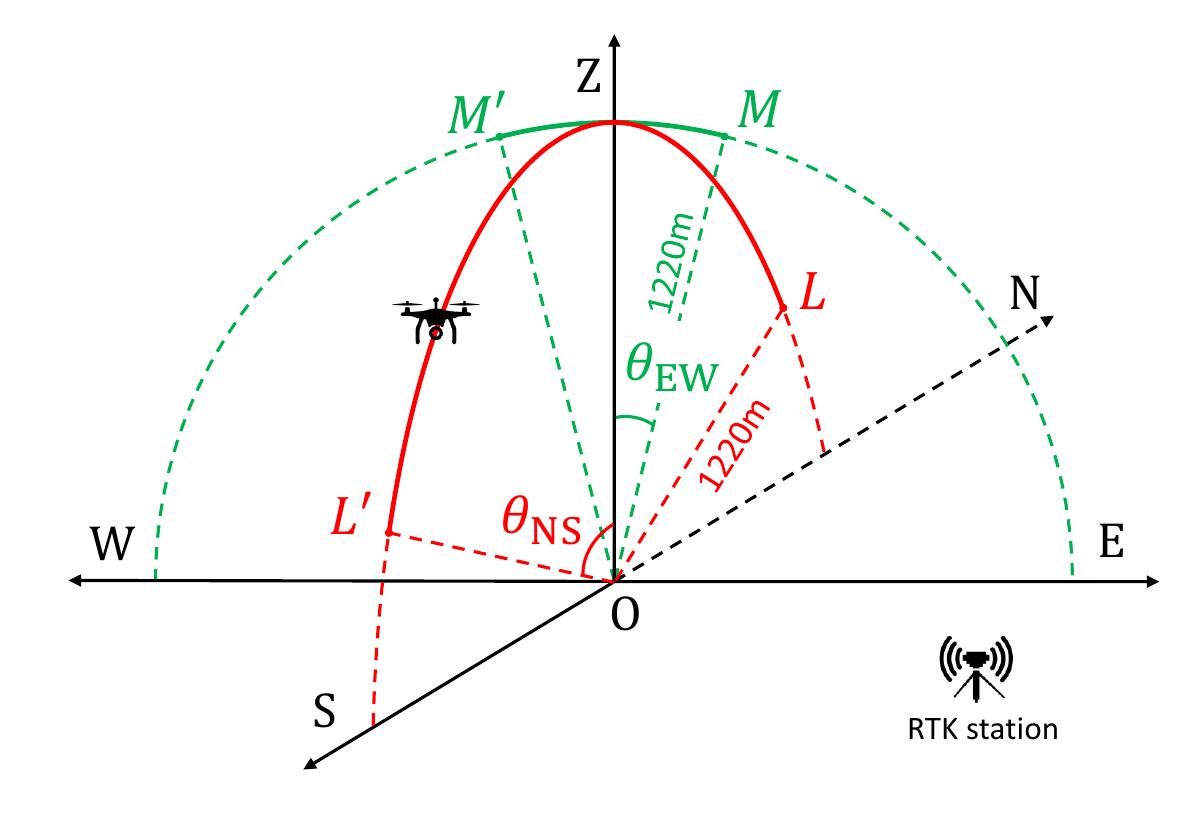}
	\caption{A schematic of the flight routes. Curve $\widearc{LL'}$ and $\widearc{MM'}$ are the circular routes centered by one feed in the origin $O$ with radius of 1220 m. The angle range in N-S direction is $-70^\circ < \theta_\mathrm{NS} < 70^\circ$. The angle range in E-W direction is $-3^\circ < \theta_\mathrm{EW} < 3^\circ$.}
    \label{fig:flight_route}
\end{figure}

Ideally, one could measure the antenna beam pattern by taking  many different points on a spherical surface centered on one of the antenna feeds. However, this would take a very large amount of time. In a simplified treatment, we may regard the cylinder beam as a product of its central pattern along the N-S direction and the pattern in the E-W direction, as assumed in Eq.~(\ref{eq:decomp}). We then choose two circular flight paths centered around a particular feed with a radius of 1220~m. A number of way points along this circular curve is preset. At each point the drone will hover for 20 seconds, with its noise source broadcasting and TCPA receiving. The flight plan is illustrated in Fig.~\ref{fig:flight_route}.
Here, the feed is at the Origin marked by $O$. In the N-S flight, the curve $\widearc{LL'}$ is the flight route. The zenith angle $\theta_\mathrm{NS}$ refers to the angle between the drone as seen by the measured feed and the zenith direction, that is $\theta_\mathrm{NS} = \angle L'OZ$ when the drone is at the point $L'$. 
We define $\theta_\mathrm{NS}$ to be negative when the drone is in the northern part and positive in the southern part. Similarly, in the E-W flight we define the angle $\theta_\mathrm{EW}$ to be positive in the eastern part and negative in the western part. As the Tianlai primary beam is narrow in the E-W direction and wide in the N-S direction, the maximum zenith angle $\theta_\mathrm{NS}$ in N-S direction we measure is $\pm70^\circ$, and $\theta_\mathrm{EW}$ in E-W direction $\pm3.0^\circ$.

The dipole antenna for the noise emitter carried by the drone produces a linearly polarized signal along the direction of the dipole, and only the corresponding polarization of the cylinder feed is sensitive to its emission. In order to measure the response for both polarizations of the feeds, two flights are needed for each path, with the dipoles pointing to the N-S direction and E-W direction once each.

For the E-W flight, the main beam ($-3^\circ$ to $3^\circ$) is very narrow, so a step size of $0.3^\circ$ is used.
For the N-S flight, since the beam width is very wide, larger zenith angle steps of $5^\circ, 10^\circ$ and even $15^\circ$ are used. These zenith angles of the measurements for the two flight curves are listed Table~ \ref{tab:zenith_angles}. 

We use the  the drone auto-pilot to execute all flights. Nevertheless, strong wind warning may interrupt the process, so manual control to resume the automatic pilot is occasionally needed. Also, when the drone is in the center line of the beam where the antenna gains are the strongest, it is necessary to decrease the level of the noise calibrator emission power to avoid saturating the receiver on the TCPA. This is executed by adjusting the attenuator, with the attenuation factor divided in the data processing to compensate for this decrease of power.

The Sun can generate a strong signal at this band throughout the daytime which can bias the measured signal level. To avoid this, the experiment is carried out at night. This has the additional advantage that at night, the ambient temperature is more stable, which yields a more stable system gain, especially after midnight. Also, the moment when transits of the bright radio sources such as Cyg A and Cas A are avoided, to ensure the drone is the only bright source in the beam when the measurement is taken. 
Our experiment were done between midnight and dawn before sunrise, when the temperature variation is the least. The system gain can be regarded as almost constant during each 15 minute flight.

\begin{table*}
    \centering
    \caption{Zenith angle steps for the drone in N-S direction and E-W direction.}
    \label{tab:zenith_angles}
    \begin{tabular}{l|l}
    \hline
      Beam part & Zenith angles (degree) \\
    \hline
     $\theta_{\rm NS}$  & -70, -55, -40, -30, -20, -10, -5, 0, +5, +10, +20, +30, +40, +55, +70 \\
     $\theta_{\rm EW}$  & -3.0, -2.7, -2.4, ..., -0.3, 0.0, +0.3, ..., +3.0 \\
      \hline
    \end{tabular}
\end{table*}

\subsection{Data Processing Procedure}
\label{sec:data_prepare}

We use the TCPA correlator to record the data. This correlator is based on a Roach+GPU framework and can compute visibilities from all of the 192 polarized, i.e $(31+32+33) \times 2$  signals of the cylinder array,  in the frequency range of 700-800~MHz with a resolution of 122~kHz \citep{Wangzh2024roach}, and 4 seconds visibility integration time.  For the 20 seconds hovering time of the drone, about 4 data points for each zenith angle are obtained, from which we calculate the average value of the auto-correlations $V^{ii}$ for each of the 192 signals. The observed visibilities can be written as (here we omit the subscripts denoting the array element pair)
$$V_{\rm{obs}}(\nu,t) = V_{\rm{dns}}(\nu,t) + V_0(\nu,t) ,$$ 
where $V_{\rm{obs}}(\nu,t)$ is the observed output from the correlator, $V_{\rm{dns}}(\nu,t)$ is the contribution of the drone noise source, and $V_0(\nu,t)$ is the visibility output without the drone, which includes the sky signal and the receiver noise. The noise source on the drone is set to a sufficiently high power level for better signal to noise ratio, but still within the range of linearity of the system. The visibility of the drone noise source can then be obtained by
\begin{equation}
V_{\rm dns}(\nu, t)= V_{\rm obs}(\nu, t)-V_0(\nu, t)
\end{equation}

\begin{figure}
    \centering
    \includegraphics[width=\textwidth]{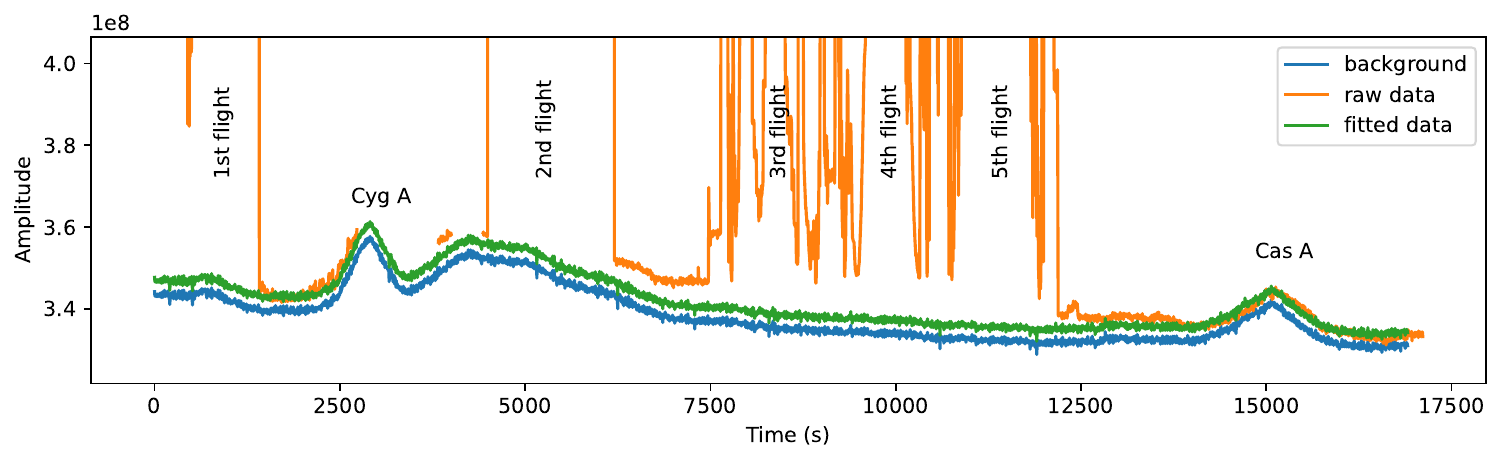}
    \caption{Fitting background data from adjacent days to data of the drone experiment. Curves here are the auto-correlation of the E plane for feed A25 XX polarization. Orange curve: the raw data observed in the whole night of about 5 hours when 5 flights are accomplished. Blue curve: the background sky amplitude taken during an adjacent day during the same sidereal time. Green curve: the linearly fitted data using the background sky data. The time periods of (1616s to 2079s) and (13356s to 16654s), when the noise source on the drone is powered off, are used for the fitting.}
    \label{fig:fit_sky}
\end{figure}

If the noise source could be switched on and off during the calibration flight, then it would be relatively straightforward to measure $V_{\rm obs}$ and $V_0$ as the on and off values. Unfortunately, during the experiment this is not done. The drone continuously broadcasts its noise signal.  Instead, we use the data taken one day before the drone experiment (July 7th, 2024), as the baseline data to derive $V_0$, while the data taken on July 8th and 9th, 2024, during which the drone experiments were performed, as $V_{\rm obs}$. However, the gain of the receiver may also drift during the different days and flights. To address this problem, we model the gain as varying according to $g(\nu,t)=(a \cdot t +b)$; i.e. it is frequency independent and varies linearly with time. 

In Fig. \ref{fig:fit_sky}, we show the visibility data for feed A25 XX polarization at frequency 720~MHz. The background sky data (from July 7th 2024)  is shown as the blue curve. The curve shows a continuous and slowly varying 
sky and receiver noise signal, over about 5 hours during the night. The peaks or increase of the signal level
 during the transits of Cyg A (at time offset of about 3000 seconds), Milky Way (at time offset of about 4500 seconds) and Cas A (at time offset of about 15000 seconds) can be clearly seen. 
 The orange curve shows the data during 5 flight sessions during the experiment of July 8th, 2024. The noise source signal level from the drone is much higher than the background signal, about 10-15 times stronger at the center of the beam and is shown truncated in this plot. Note that for the orange curve, when Cyg A transits the antenna beam, we paused the drone experiment and the noise source is kept powered off. Visibility data from two time slices, (1616 s to 2079 s) and  (13356 s to 16654 s)  are used to linearly fit the background sky curve with the assumption that the gain varies linearly with time and independent of frequency, as noted above, so that 
$$V_\mathrm{fit}(\nu, t) = (a\cdot t + b)V_\mathrm{sky}(\nu, t),$$ 
The green curve shows the fitted sky background. Finally, we obtain the sky removed visibility,
\begin{equation}
V_\mathrm{dns}(\nu, t) = V_\mathrm{obs}(\nu, t) - V_\mathrm{fit}(\nu, t)
\end{equation}
For the beam pattern determination analysis, we convert
 $V_{\rm dns}(\nu, t)$ to $\hat{V}_{\rm dns}(\nu, \theta)$ by averaging the visibility over the time when the drone hovered for 20 seconds.

\section{Calibration of the Drone Emitter}
\label{sec:drone_cali}
\begin{figure}
    \centering
    \includegraphics[width=0.4\textwidth]{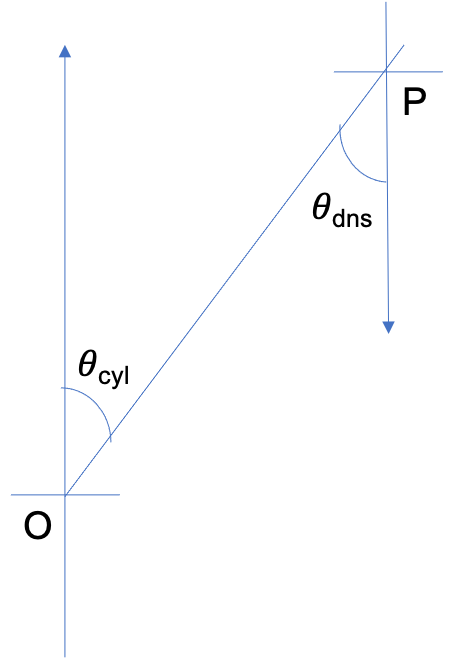}
    \includegraphics[width=0.4\textwidth]{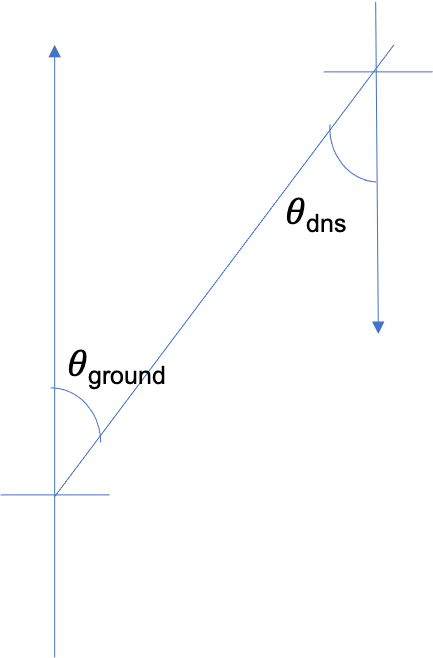}
    \caption{Left: The signal received by the cylinder from the drone depends; Right: The signal received by the ground dipole for the calibration of the drone beam.}
    \label{fig:drone_calib}
\end{figure}

The measured signal strength is proportional to the beam of the cylinder antenna in the direction of the drone with respect to the cylinder, but it is also proportional to the beam of the noise emitter antenna in the direction of the cylinder with respect to the drone.
As shown the left panel of Fig.~\ref{fig:drone_calib},
the received signal strength is 
\begin{eqnarray}
 V_{\rm dns} &\propto&  B_{\rm cyl}(\vec{OP}) 
\cdot B_{\rm dns}(\vec{PO})\nonumber\\
&=&B_{\rm cyl}(\theta_{\rm cyl})
\cdot B_{\rm dns}(\theta_{\rm dns}).
\label{eq:dns}
\end{eqnarray}
where the vector $\vec{OP} $ and $\vec{PO}$ can be marked by the angle $\theta_{\rm cyl} $ and $\theta_{\rm dns}$.
Thus, to measure the cylinder beam data using the drone, we first need to know the beam pattern of the antenna mounted on the drone (hereafter called drone beam $B_{\rm dns}$). 
While the beam pattern of a dipole antenna is well known, the presence of the drone can affect this beam. It is therefore necessary to first measure the drone beam. The drone beam is in turn measured by a dipole antenna on the ground (hereafter its beam is called ground dipole beam $B_{\rm ground}$), mounted about 10~cm above the ground surface during the measurement, as shown in the right panel of Fig.~\ref{fig:drone_calib}. In this case, we would have
\begin{equation}
V_{\rm dns} \propto B_{\rm ground}(\theta_{\rm ground})
\cdot B_{\rm dns}(\theta_{\rm dns}).
\label{eq:ground}
\end{equation}
We will first use Eq.~(\ref{eq:ground}) to measure $B_{\rm dns}(\theta_{\rm dns})$, assuming that the beam of the ground dipole $B_{\rm ground}(\theta_{\rm ground}) $ is known, 
then use Eq.~(\ref{eq:dns}) to measure $ B_{\rm cyl}(\theta_{\rm cyl})$.

\subsection{The Beam of the Ground Dipole}
\label{sec:ground_beam}

We use the \texttt{FEKO} electromagnetic (EM) simulation software to simulate the E plane (parallel to the dipole) of the ground dipole antenna. 
The relative permittivity or dielectric constant of the ground is set to $\epsilon_{Gnd}/\epsilon_0 \simeq 3.66$ and its conductivity $\sigma_{Gnd} \simeq 0.082 \mathrm{S/m}$, corresponding to the measured values, although not at the exact position of the antenna during the measurements.
These parameters are measured using the standard four probe method (\citealt{dielectric}, see also \citealt{Suonanben2024}.
The ground is set to be infinitely large. In the real case, the experiment ground is nearly flat and has no obstructions over 200 meters.

\begin{figure}
    \centering
    \includegraphics[width=0.7\linewidth]{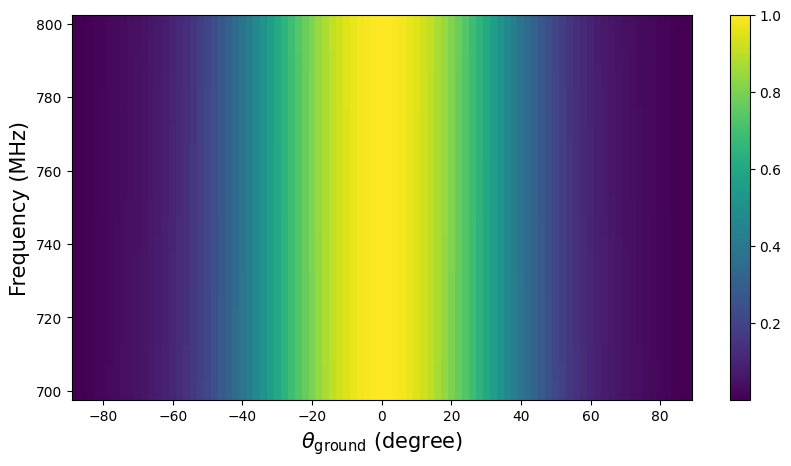}
    \caption{The simulated beam of a dipole antenna (E-plane) located 10~cm above the ground with relative dielectric constant of 3.66 and conductivity of 0.082 S/m. }
    \label{fig:groundbeam_sim}
\end{figure}

In Fig. \ref{fig:groundbeam_sim}, we show the simulated beam in the E plane (top panel) of the ground dipole antenna beam. The beam is symmetric about a zenith angle of 0 degree. The beam can be approximated as a Gaussian function of zenith angle. At the high frequency end, the beam width is slightly narrower than the low frequency end, as expected.

\begin{figure*}
    \centering
    \includegraphics[width=0.8\linewidth]{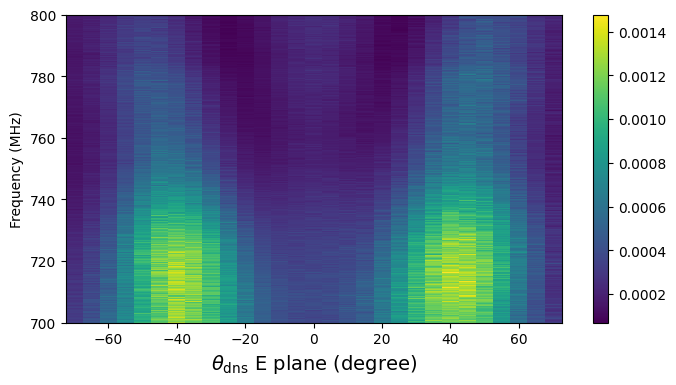}\\
    \includegraphics[width=0.8\linewidth]{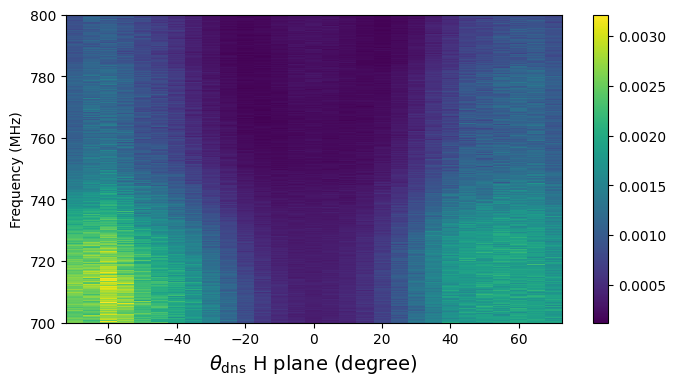}
    \caption{The measured drone beam after correction with the beam pattern of the ground dipole antenna 
    for the E-plane (top) and the H-plane (bottom).}
    \label{fig:drone_beam}
\end{figure*}

\subsection{The Beam of the Drone Dipole}
\label{sec:drone_beam}

In the experiment to measure the drone beam pattern $B_{\rm dns}$, the drone flies from one side to the other side of the ground dipole in an arc with radius of 100~meters, and hovers at zenith angles from $-75^\circ$ to $75^\circ$ with a step of $5^\circ$.
Both the E and H planes are measured. In the E plane flight, the dipole direction is the same as the flight route direction, while in the H-plane, the flight route directions are perpendicular to that of the dipole. The polarization of the ground dipole is always the same as the drone dipole.  
The ground dipole antenna receives the signal from the drone dipole. A spectrum analyzer is used to measure and record the received signal between 700-800~MHz.
Hereafter the raw drone beam is called $B_{\rm dns\_raw}$, so that $B_{\rm dns\_raw} \propto B_{\rm dns} \cdot B_{\rm ground}$. Then, we can obtain the real drone beam after calibrating by the ground dipole antenna beam,
\begin{equation}
B_{\rm dns}(\nu, \theta) \propto \frac{ B_{\rm dns\_raw}(\nu, \theta)}{ B_{\rm ground}(\nu, \theta)}
\label{eq:B_dns}
\end{equation}

Figure \ref{fig:drone_beam} shows the E plane (top) and the H plane (bottom) of the drone beam $B_{\rm dns}$, after correct for the ground dipole profile as in Eq.~(\ref{eq:B_dns}). As can be seen from this figure, the beam pattern of the dipole antenna mounted on the drone is different from that in free space, for example it has two lobes in the downward direction, probably caused by the induction and reflections from the main frame of the drone. 

\begin{figure}
    \centering
    \includegraphics[width=0.6\linewidth]{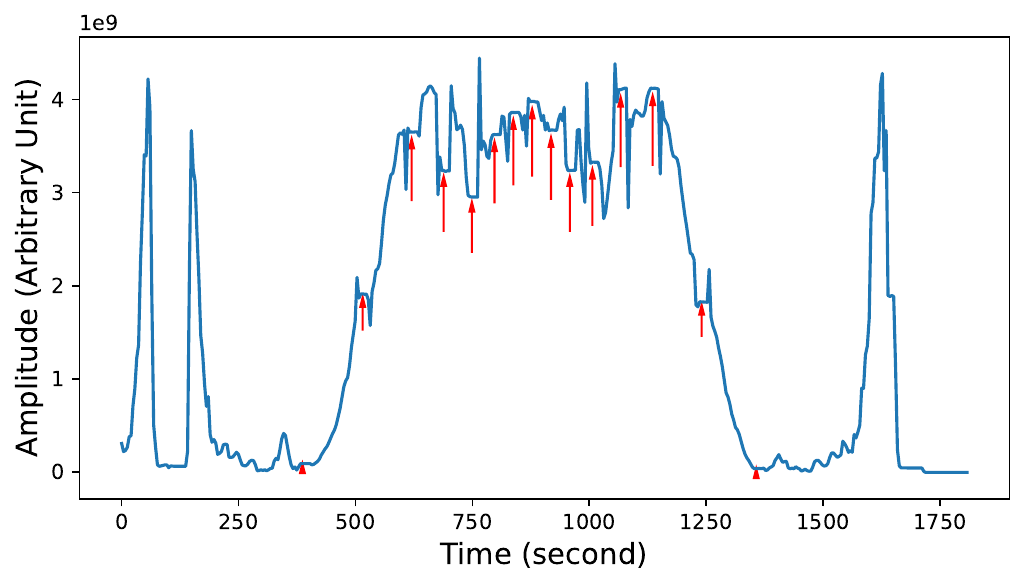}\\
    \includegraphics[width=0.6\linewidth]{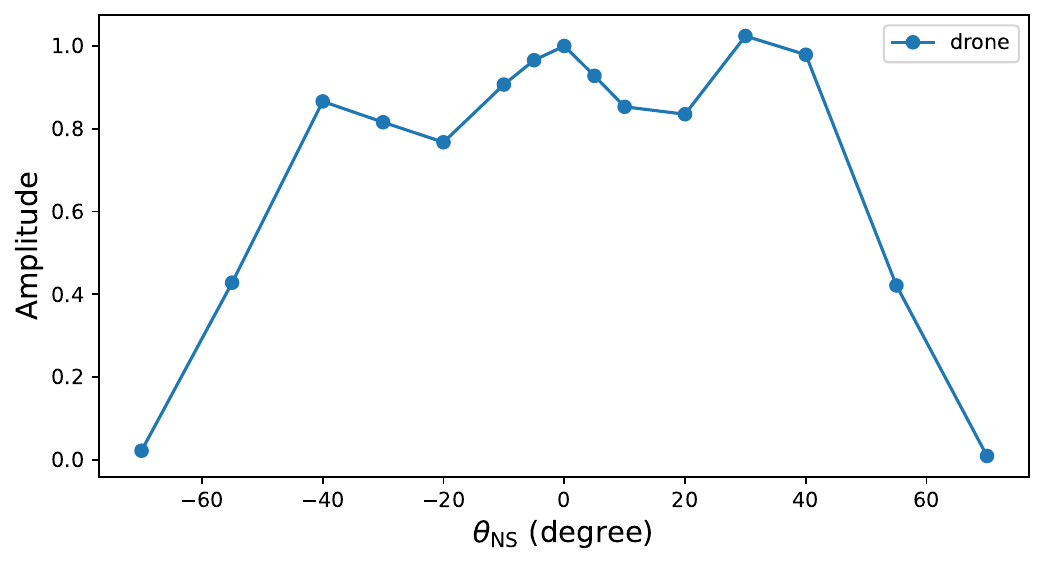}\\
    \caption{Top: the raw amplitude of the A25 XX polarization as a function of time after the drone took off.  Red arrows indicate the time for the drone hovering at the preset way points.
    Bottom: average amplitudes of the signal as a function of zenith angle $\theta_\mathrm{NS}$.
    }
    \label{fig:processing}
\end{figure}

\section{Measurement Results of the Cylinder Beam}
\label{sec:results} 

Having measured the drone beam needed for making corrections, we can now measure the beam of the cylinder. Taking the E plane measurement for the XX polarization of feed A25 as an example.
First, the auto-correlation of the designated channel is extracted from the data file using the {\tt tlpipe} software \citep{Zuosf2021tlpipe}. The sky background is removed using the adjacent day data as discussed in Sec. \ref{sec:data_prepare}. 
The time-ordered data is shown in the top panel of Fig. \ref{fig:processing}. 
The time when the drone is hovering at the way points can be recognized, these locations are then transformed to zenith angles $\theta$ of the drone relative to feed A25, and the amplitude of the received signal is obtained by averaging over the duration of the hovering, as shown in the bottom  panel of Fig.~\ref{fig:processing}. 
This sky-removed visibility of the drone $\hat{V}_{\rm dns}(\nu, \theta)$ is 
\begin{equation}
    \hat{V}_{\rm dns}(\nu, \theta) \propto B_{\rm dns}(\nu, \theta) \cdot B_{\rm cyl}(\nu, \theta)
\label{eq:V_dns}
\end{equation}
where $B_{\rm cyl}$ is the cylinder beam that we want to measure, $B_{\rm dns}$ is the drone beam obtained in Eq.~(\ref{eq:B_dns}).
Using Eq.~(\ref{eq:V_dns}), the cylinder beam $B_{\rm cyl}$ is obtained by dividing the drone beam measured in Sec. \ref{sec:drone_beam} from the visibility $\hat{V}_{\rm dns}$.

\begin{figure}
    \centering
    \includegraphics[width=0.8\textwidth]{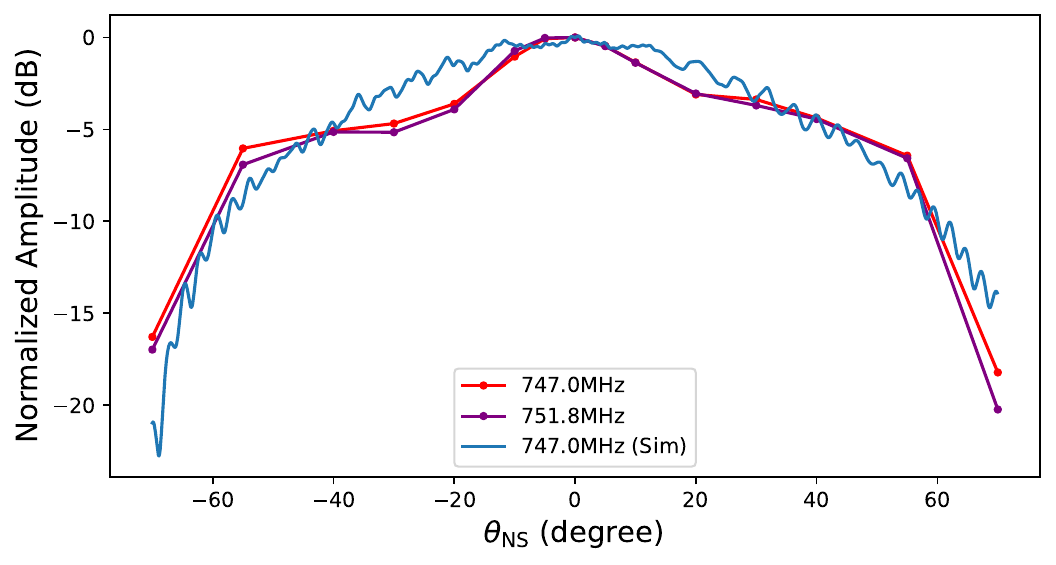}
    \caption{The measured cylinder N-S beam profile of the feed A25 XX polarization at two central frequencies.  }
    \label{fig:NS-profile}
\end{figure}

The final result of the measurement is shown in Fig. \ref{fig:NS-profile} for the N-S profile of a feed (A25 XX), where the amplitude has been re-normalized at $\theta_\mathrm{NS} = 0^\circ$, and shown in logarithmic (dB) scale. The results for two frequencies ( 747.0 MHz and 751.8 MHz), averaged in frequency over 0.98 MHz are plotted, showing similar beam profiles for these two neighbouring frequencies.  
From Fig.\ref{fig:NS-profile}, we see this N-S beam profile of the cylinder has a peak at $\theta_\mathrm{NS} = 0^\circ$, as we move away from the peak to the two sides, the beam declines, first to a `shoulder' at $\theta_\mathrm{NS} \sim \pm 30^\circ $, then steepens at $\theta_\mathrm{NS} \sim \pm 60^\circ$. The shape is slightly asymmetric, with the `shoulder' more apparent on the $\theta_\mathrm{NS} < 0$ side, i.e. the northern part of N-S beam. 
We also plotted the simulation result on the same figure for comparison, which does not have any shoulder, with a symmetric profile and monotonous decline from the peak. Thus, at the shoulder the difference can be as large as 3 dB.

\begin{figure}
    \centering
    \includegraphics[width=0.8\textwidth]{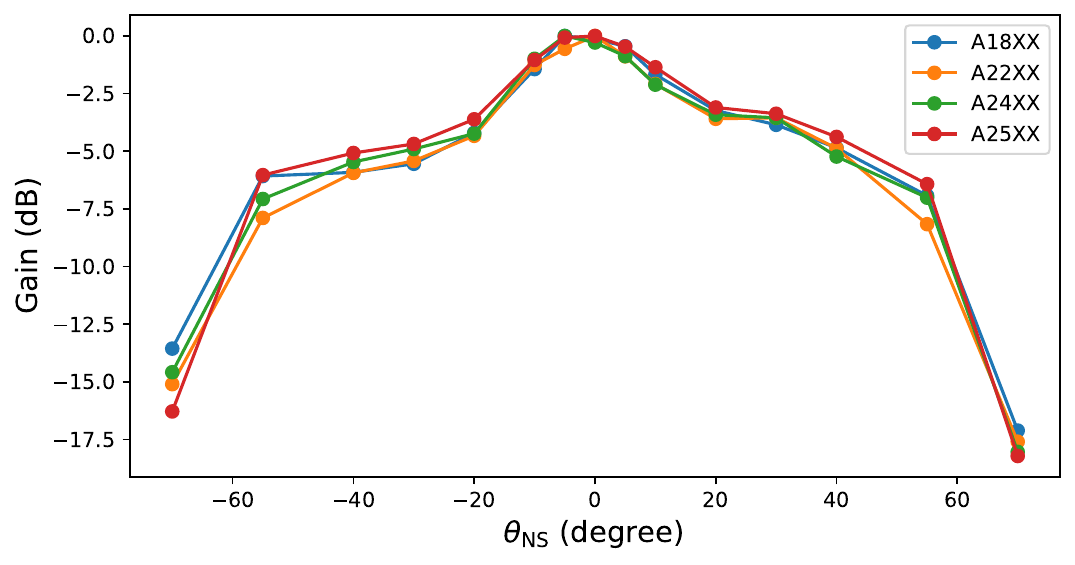}
    \caption{The N-S beam profiles of 4 feeds using the XX polarization.}
    \label{fig:cmpr_feeds}
\end{figure}

\begin{figure}
    \centering
    \includegraphics[width=0.8\textwidth]{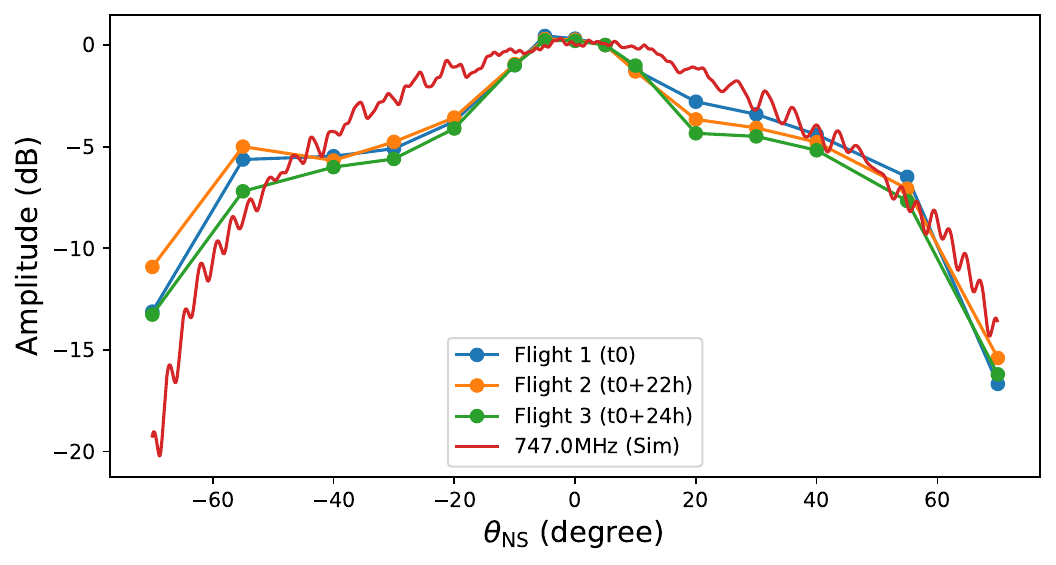}
    \caption{The N-S beam profile for Feed A18 XX obtained in three flights. }
    \label{fig:NS-profile-3flights}
\end{figure}

The measured beams for different feeds are very similar with each other. In Fig. \ref{fig:cmpr_feeds}, we show the N-S beam profiles of the XX polarization for 4 different feeds, which have very similar profiles. 
The results from different flights are also generally in agreement with each other, showing that the result of the measurement is robust. As an example, in the bottom panel of Fig. \ref{fig:NS-profile-3flights}, we plot the N-S beam profile of feed A18 XX polarization  using the data from three flights. The second flight took place 22 hours after the first flight, and the third flight took place 2 hours after the second flight, i.e. 24 hours after the first flight.  The results are consistent with each other, and also similar to that of the A25 XX.
In these measurements, the difference between the feeds are comparable to the difference in the measurement result for the same feed from different flights. 

\begin{figure}
    \centering
    \includegraphics[width=0.8\linewidth]{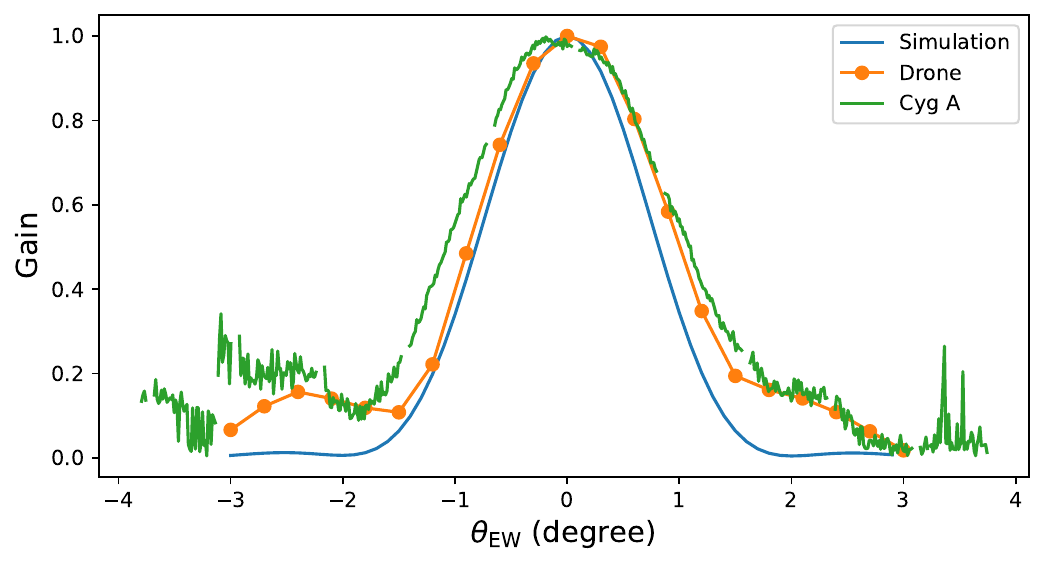}
	\caption{The E-W profile for feed A26 XX polarization, with drone measurement, Cyg A transit measurement, and EM simulation. }
    \label{fig:EW-profile}
\end{figure}

For the cause of the asymmetry seen in the measurement result, we considered several possibilities. First, there is the measurement error in the drone. If the drone is slightly tilted during its flight due to wind, or if its beam profile has some error, it could induce an asymmetry. However, this result is relatively stable over different measurements (as discussed below with Fig.\ref{fig:NS-profile-3flights}), which suggests that this may have more to do with the cylinder rather than with the drone. It could be due to slightly error in the installation of the reflector or the feed, but based on the theodolite measurements, this error should be much smaller. It seems to us that the most likely cause is ground reflections from the nearby hills. In the simulation, the cylinder antenna is placed on flat land, but in reality there are hills to the north and south of the cylinders. The asymmetric landscape may be the  cause of the observed differences in the north and south side, though this still needs to be by further experiments and analysis in the future. 

The simulation result has some kind of `jitter' with the angle due to the cross-coupling between the different feeds, which varies with both frequency and angle. This is not apparent in the measurement result, mainly because the sampling over the angles are not sufficiently dense. The number of measurement points are limited by the duration of the flight. In future works, the sampling could be made denser by dividing the measurement path into several parts, and each time complete one part with more points.

The E-W direction beam profile measured by the drone is shown in Fig. \ref{fig:EW-profile}, where the simulation result and the beam profiled measured by the transit of bright source are also shown for comparison. Here we again see that the overall shapes of the simulation are similar, but there are subtle differences between the actual measurement and the simulation. Both the drone measurement and the source transit measurement yield beam profiles which are slightly asymmetric, slightly broader than the simulation result, and higher side lobes. The drone measurement is somewhere between the simulation and the point source transit measurement result. 

Note that these results are obtained after subtracting the noise background, and away from the peak, when the noise is dominant, the inaccurate estimate of the signal may lead to apparently broader beam profile. As the drone measurements generally have a higher signal-to-noise ratio than the transit measurements, it may suffer less from this effect, and thus appear to match better with the simulation.  In addition, the surface errors can affect the broadening of the measured beam. Another factor which may lead to apparently broader profile is the finite size of the bright source, in this case the Cyg A. It is well known that Cyg A is not a point source, but has a size about 2 arcmin (the separation of its two radio lobes) \citep{Carilli96}, while the primary beam of the Tianlai cylinder is about $2^\circ$, so the size of Cyg A is too small to affect this measurement.

\begin{figure}
    \centering
    \includegraphics[width=0.49\linewidth]{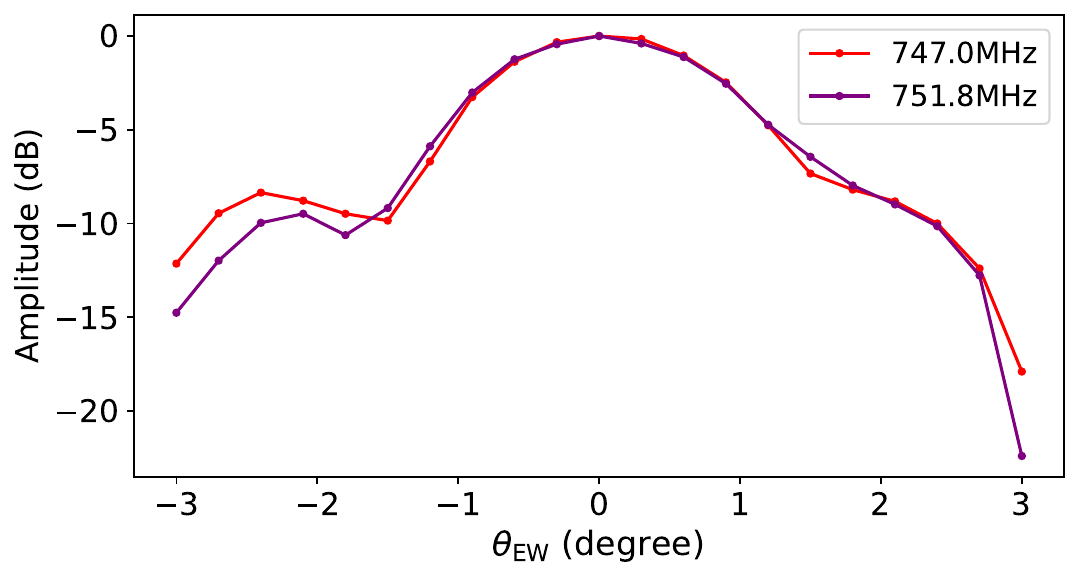}
    \includegraphics[width=0.49\linewidth]{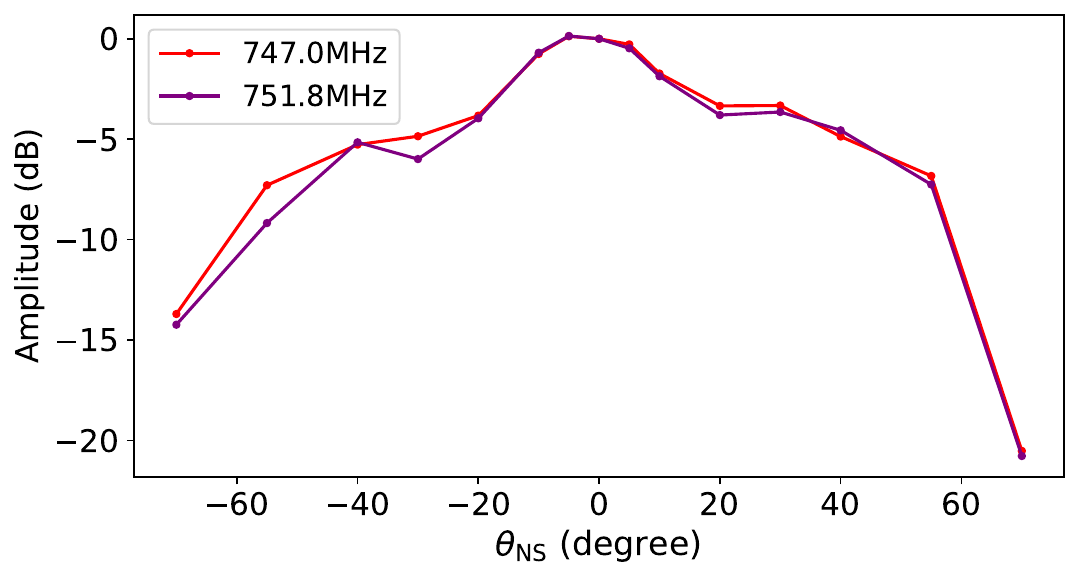} \\
    \includegraphics[width=0.49\linewidth]{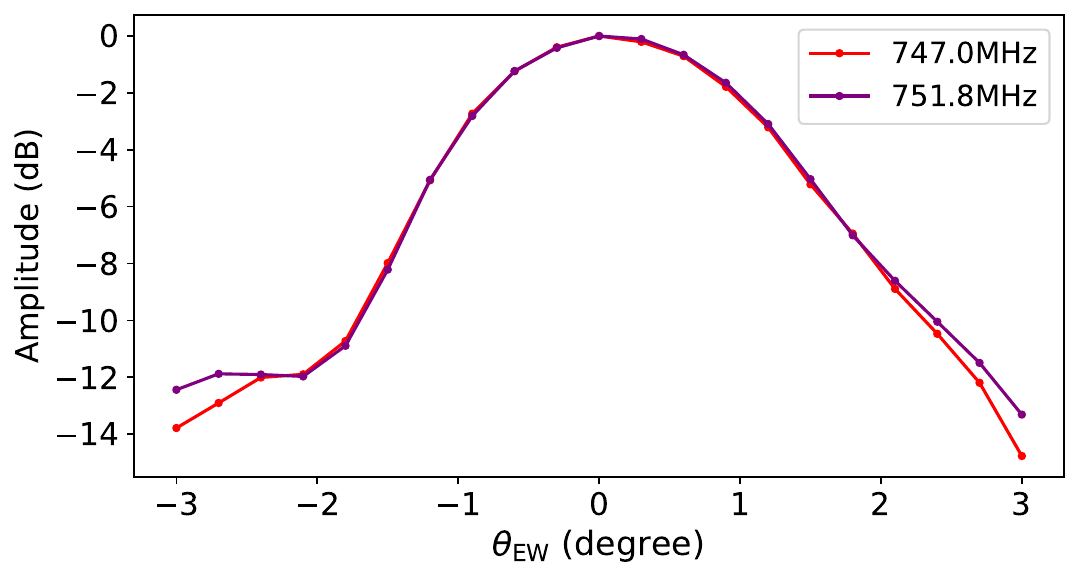}
    \includegraphics[width=0.49\linewidth]{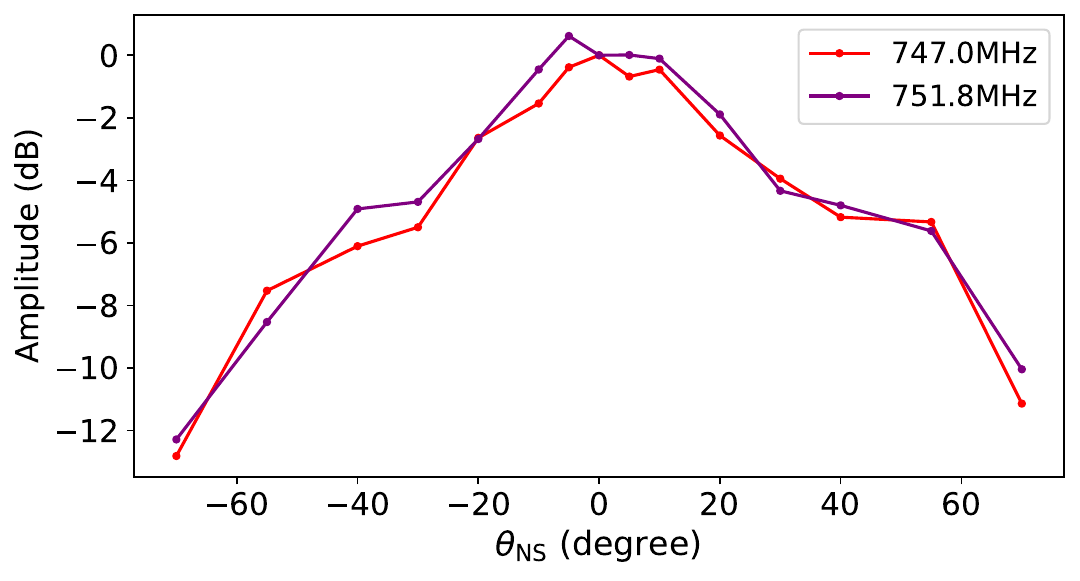}
    \caption{
     The beam profiles of E-W (left) and N-S (right) direction for feed A26 XX (top) and YY (bottom) polarization.}
    \label{fig:result_mainbeam}
\end{figure}

Finally, as yet another feed example, we show the main beams for feed A26 in both E and H planes and in XX (NS) and YY (EW) polarizations in Fig.~\ref{fig:result_mainbeam}, at 747.0 MHz and 751.8 MHz. The XX polarization result is again similar to the prevous example, but in the case of YY polarization H-plane, the peak of the profile is not at $\theta=0^\circ$, the beam has a small dip near the center $\theta=0^\circ$, which may be due to some cross-coupling effect. For the E-W profile, we also find the first sidelobes are not completely symmetric. This may be caused by the different reflections in the east and west sides as cylinder A is located in the east side of the array, and to its east there is also the dish array. These factors may affect the side lobe of the beam in the E-W direction.

\section{Conclusion}
\label{sec:discussion}

Measuring the beam pattern of a radio telescope antenna is an important part of telescope calibration. In many modern 21~cm experiments, the telescope antenna is fixed on the ground or can only be moved between observations. In these cases only the east-west direction beam can be measured conveniently by observing the meridian transit of astronomical sources, while measuring the North-South beam profile poses a significant challenge. The TCPA is an example of these. The use of UAV offers a simple and efficient means to determine the beam profile of such telescope antennas, especially along the North-South direction. 

In this paper, we detail a UAV-based far-field beam measurement method, where the North-South beam patterns were measured by hovering the drone at specific points. By comparing East-West beam measurements with astronomical observations, the UAV measurement methods and techniques were validated. This approach can be applied to various antenna beam measurements and array beam characterizations, contributing to advancements in precision radio astronomy and 21~cm intensity mapping surveys.
We find the beam profiles obtained from the measurement is generally similar to the EM simulation. However, differences do exist in the `shoulder`, sidelobes and beam width, because there are simplifying assumptions and treatments in the  simulation, which does not capture all aspects of the actual instrument in every detail, e.g. the reflections from the surrounding terrain and objects, surface imperfection, and cross-coupling noise in the reflector.

\begin{acknowledgements}
We acknowledge the support by National SKA Program of China (Nos. 2022SKA0110100 and 2022SKA0110101), the National Natural Science Foundation of China (12203061, 12361141814, 12303004, 12273070), and the Chinese Academy of Science ZDKYYQ20200008.

\end{acknowledgements}

\bibliographystyle{raa}
\bibliography{refs}

\label{lastpage}

\end{document}